\documentclass[sn-mathphys,Numbered]{sn-jnl} 
\usepackage{fix-cm} 

\usepackage{multirow}%
\usepackage{amsmath,amssymb,amsfonts}%
\usepackage[title]{appendix}%
\usepackage{xcolor}%
\usepackage{manyfoot}%
\usepackage{booktabs}%
\usepackage{longtable}%

\begin{document}

\title[Ariel de-jittering]{De-jittering Ariel: an optimized algorithm}


\author*[1,2]{\fnm{Andrea} \sur{Bocchieri}}\email{andrea.bocchieri@uniroma1.it}

\author[3]{\fnm{Lorenzo V.} \sur{Mugnai}}

\author[1]{\fnm{Enzo} \sur{Pascale}}

\author[3]{\fnm{Andreas} \sur{Papageorgiou}}

\author[4]{\fnm{Angèle} \sur{Syty}}

\author[5]{\fnm{Angelos} \sur{Tsiaras}}

\author[6]{\fnm{Paul} \sur{Eccleston}}

\author[7]{\fnm{Giorgio} \sur{Savini}}

\author[7]{\fnm{Giovanna} \sur{Tinetti}}

\author[8]{\fnm{Renaud} \sur{Broquet}}

\author[9]{\fnm{Patrick} \sur{Chapman}}

\author[10]{\fnm{Gianfranco} \sur{Sechi}}



\affil*[1]{\orgdiv{Dipartimento di Fisica}, \orgname{La Sapienza Università di Roma}, \orgaddress{\street{Piazzale Aldo Moro 5}, \city{Roma}, \postcode{00185}, \country{Italy}}}

\affil*[2]{\orgdiv{INAF}, \orgname{Osservatorio Astrofisico di Arcetri}, \orgaddress{\street{Largo Enrico Fermi 5}, \city{Firenze}, \postcode{50125}, \country{Italy}}}

\affil[3]{\orgdiv{School of Physics and Astronomy}, \orgname{Cardiff University}, \orgaddress{\street{Queens Buildings, The Parade}, \city{Cardiff}, \postcode{CF24 3AA}, \country{UK}}}

\affil[4]{\orgdiv{Institut d’Astrophysique de Paris}, \orgname{Sorbonne Université, CNRS}, \orgaddress{\street{98 bis bd Arago}, \city{Paris}, \postcode{75014}, \country{France}}}

\affil[5]{\orgdiv{Department of Physics}, \orgname{Aristotle University of Thessaloniki}, \city{Thessaloniki}, \postcode{54124}, \country{Greece}}

\affil[6]{\orgname{RAL Space}, \orgname{STFC}, \orgaddress{\street{Fermi Ave}, \city{Harwell}, \postcode{OX11 0QX}, \country{UK}}}

\affil[7]{\orgdiv{Department of Physics and Astronomy}, \orgname{University College London}, \orgaddress{\street{Gower Street}, \city{London}, \postcode{WC1E 6BT}, \country{UK}}}

\affil[8]{\orgname{Airbus Defence \& Space}, \orgaddress{\street{Rue des Cosmonautes 31}, \city{Toulouse}, \postcode{31400}, \country{France}}}

\affil[9]{\orgname{Airbus Defence \& Space}, \orgaddress{\street{Gunnels Wood Road}, \city{Stevenage}, \postcode{SG1 2AS}, \country{UK}}}

\affil[10]{\orgname{European Space Research \& Technology Centre}, \orgaddress{\street{Keplerlaan 1}, \city{Noordwijk}, \postcode{2200 AG}, \country{The Netherlands}}}

\abstract{
The European Space Agency's \textit{Ariel} mission, scheduled for launch in 2029, aims to conduct the first large-scale survey of atmospheric spectra of transiting exoplanets. \textit{Ariel} achieves the high photometric stability on transit timescales required to detect the spectroscopic signatures of chemical elements with a payload design optimized for transit photometry that either eliminates known systematics or allows for their removal during data processing without significantly degrading or biasing the detection.
Jitter in the spacecraft's line of sight is a source of disturbance when measuring the spectra of exoplanet atmospheres. We describe an improved algorithm for de-jittering \textit{Ariel} observations simulated in the time domain. We opt for an approach based on the spatial information on the Point Spread Function (PSF) distortion from jitter to detrend the optical signals.
The jitter model is based on representative simulations from Airbus Defence and Space, the prime contractor for the \textit{Ariel} service module. We investigate the precision and biases of the retrieved atmospheric spectra from the jitter-detrended observations. 
At long wavelengths, the photometric stability of the \textit{Ariel} spectrometer is already dominated by photon noise. Our algorithm effectively de-jitters both photometric and spectroscopic data, ensuring that the performance remains photon noise-limited across the entire \textit{Ariel} spectrum, fully compliant with mission requirements. 
This work contributes to the development of the data reduction pipeline for \textit{Ariel}, aligning with its scientific goals, and may also benefit other astronomical telescopes and instrumentation.
}


\maketitle

\section{Introduction}\label{sec:introduction}

\textit{Ariel} ~\citep{Tinetti:2018,Tinetti:2021} is the fourth medium-class mission of the European Space Agency (ESA) Cosmic Vision programme, scheduled for launch in 2029.
The mission will perform the first spectroscopic survey of the atmospheres of a large and diverse sample of exoplanets, enabling comparative studies of the physics and chemistry of exoplanets. 
The observations, covering the visible to infrared spectral range (0.5--7.8 \textmu m), will be conducted using a 1-meter class Cassegrain telescope~\citep{Pace:2022}.

The payload comprises two instruments: the Fine Guidance System (FGS)~\citep{Rataj:2019} and the \textit{Ariel} Infra-Red Spectrometer (AIRS)~\citep{Martignac:2022}. 
FGS has three photometers (VISPhot, 0.50 -- 0.60 \textmu m; FGS-1, 0.60 -- 0.80 \textmu m; FGS-2, 0.80 -- 1.10 \textmu m) and one spectrometer (NIRSpec, 1.10 -- 1.95 \textmu m and $R \geq 15$); AIRS has two spectrometers (AIRS-Ch0, 1.95 -- 3.90 \textmu m and $R \geq 100$;  AIRS Ch1, 3.90 -- 7.80 \textmu m and $R \geq 30$). 

\textit{Ariel}'s orbit around the Sun-Earth Lagrange point L2 ensures a photometrically and thermally stable environment for the telescope and instruments. 
The required photometric stability to measure atmospheric signals from the planet is 10--100 parts per million (ppm) relative to the stellar flux \citep{Tinetti:2018}, to be achieved on the relevant timescale of transit observations. Science requirements set the typical observation duration to be 2.5 times the exoplanet transit duration, with typical observations lasting about 10 hours. While \textit{Ariel} can perform longer observations, these quantities provide useful design points, translating into transit durations of a few hours. This allows us to set the relevant timescale of transit observations at approximately 1 hour.

The photometric stability of the \textit{Ariel} payload is essential for achieving its scientific objectives. The primary requirement is that all sources of uncertainty must not significantly increase the photometric noise from the astrophysical scene (star, planet, and zodiacal light) over the typical timescales relevant for transit and eclipse events, which range from about 5 minutes to 10 hours.

The design of the \textit{Ariel} payload incorporates lessons learned from measuring exoplanetary atmospheres using Spitzer, HST, and ground-based instruments. \textit{Ariel} will collect spectroscopic and photometric time series of transiting exoplanets with a stability better than 10 to 100 ppm over a single transit observation, depending on the target's brightness. Key features that enable \textit{Ariel} to achieve this stable performance include:

\begin{enumerate}
    \item Simultaneous observations of the same transit event across all photometric and spectroscopic channels.
    \item Continuous observation of the transit event to ensure measurements are made in thermally and photometrically stable conditions.
    \item A payload design that makes \textit{Ariel} resilient to major sources of systematics or allows for their removal during post-processing.
\end{enumerate}

The most significant disturbances of astrophysical and instrumental origin are listed in Table~\ref{tab1} \citep[reproduced from][]{Tinetti:2021}, along with the strategies employed to mitigate their impact on detection and photometric stability.

\begin{table}[h]
\caption{Summary of noise sources and systematic errors}\label{tab1}%
\begin{tabular}[]{@{}lp{4cm}p{5.5cm}@{}}
\toprule
\textbf{Uncertainty type} & \textbf{Source}  & \textbf{Mitigation strategy} \\
\midrule
\multirow{3}{*}{Detector noise}   & Dark current   &  \multirow{2}{*}{Choice of low-noise, cryogenic detectors.}   \\
 &  Readout noise    \\ \cmidrule{2-3}
 & Gain stability & Temperature controlled focal planes, calibration, data-analysis.\\
 \midrule
\multirow{2}{*}{ Thermal noise}   & Emission from telescope, all optical elements and enclosures   & Negligible due to cryogenic temperatures.   \\
\cmidrule{2-3}
&Temperature fluctuations   & Negligible by design.   \\
\midrule
\multirow{3}{*}{Astrophysical noise}    & Target photon noise   & Fundamental noise limit, choice of telescope aperture size.   \\
\cmidrule{2-3}
& Photon noise from Zodiacal light and field stars   & Negligible over {\it Ariel} bands.   \\
\cmidrule{2-3}
& Target star activity   & Multi-wavelength stellar monitoring, post-processing detrending.   \\
\midrule
\multirow{2}{*}{Pointing jitter}    & AOCS, structural modes, detector intra/inter pixel response   & High AOCS pointing stability and post-processing detrending.   \\
\cmidrule{2-3}
& Slit losses   & Choice of slit-less prism spectrometers.   \\

\botrule
\end{tabular}
\end{table}

In this work, we describe the data processing of the time series necessary to correct the photometric noise induced by the jitter of the spacecraft's (S/C) Line of Sight (LoS), hereafter referred to as ``jitter.'' Jitter arises from the inevitable vibrations and drifts of the S/C in space, resulting in a quasi-harmonic signal that introduces spurious modulation into the data. This effect is analogous to taking long-exposure images with a shaky hand. If left uncorrected, it would provide excess noise above photon in the short wavelength channels.

A significant effort has been dedicated to developing a jitter-detrending algorithm capable of efficiently correcting this source of photometric uncertainty. Here, we present an improved version, building on previous iterations, and evaluate its performance using simulated \textit{Ariel} observations. If implemented in the data reduction pipeline, the algorithm is expected to be validated through testing during commissioning.

Our analysis focuses on the achieved photometric performance relative to mission requirements, as well as potential biases in the retrieved atmospheric spectra. Characterizing these biases is crucial for ensuring accurate data interpretation, making it a key aspect of this work.

We divide this paper into the following sections: Section~\ref{sec:methods} describes the data and methods utilized. 
Section~\ref{sec:methods:data-generation} describes the data generation. 
 Section~\ref{sec:methods:signal-extraction} describes how the signal is extracted and Section~\ref{sec:methods:algorithm} presents the jitter-detrending algorithm.  
Section~\ref{sec:results} discusses the results of the jitter detrending, with Section~\ref{sec:results:noise} focusing on noise performance and Section~\ref{sec:results:bias} on biases in the retrieved atmospheric spectrum of the planet. 
Finally, Section~\ref{sec:conclusions} concludes this work, summarizing the main points and providing future perspectives.

\section{Methods}\label{sec:methods}

\textit{Ariel} consists of both a ground and space segment. The space segment includes the payload and the service module. The \textit{Ariel} Mission Consortium (AMC) provides the payload, while Airbus Defence and Space (ADS) is the prime contractor for the service module (SVM). The SVM hosts the Attitude and Orbit Control System (AOCS), which relies on measuring the centroid of the stellar target, specifically the exoplanet host star imaged by FGS-2, to close the loop. FGS-1 can be utilized in the AOCS loop in the unforeseen event of a failure of FGS-2.

The control loop stabilizing the S/C in L2's low-gravity environment relies solely on reaction wheels (RWs) as actuators to meet the fine-pointing requirements. During continuous observation, the RWs accelerate to compensate for the torque the solar radiation pressure applies to the S/C structure. However, as the RWs accelerate, the frequency of the vibrations transferred to the S/C structure changes. Despite the RWs being mounted on vibration dampers to minimize the effect, the S/C's mechanical transfer function translates these vibrations, resulting in jitter with varying amplitudes at different frequencies. The mechanical cryocooler used to cool the AIRS detectors exports vibrations that also cause variations of the LoS, but at frequencies that do not change in time. 

Since before adoption, \textit{Ariel} has developed advanced simulators able to model the expected performance of the payload and complex astrophysical and instrumental systematics, including pointing jitter. 
\texttt{ExoSim2}\footnote{\url{https://exosim2-public.readthedocs.io/}}~\citep{Mugnai:2025} and its previous iteration \texttt{ExoSim}~\citep{Sarkar:2021} can simulate an end-to-end observation in the time domain, from the astrophysical source to the focal planes of the instruments. The early development of these advanced simulators was motivated by the desire to avoid unnecessary complexity in the payload design, thereby minimizing potential risks to the mission.

Payload requirements specify that the systematic component of pointing jitter must be detrended to become negligible relative to the exoplanet's photon noise, or be less than 20 ppm over a 1-hour timescale in each photometric channel and spectroscopic bin. Additionally, the instrument's noise level must be no higher than 20 ppm for all noise sources (including jitter) in each observation (noise floor). This implies that integrating observations to achieve a lower noise level below 20 ppm -- likely due to residual statistical correlations in the data that can affect measurements even after averaging -- is not required.

Previous analyses during Phases A and B of the mission (currently in C) demonstrate that \textit{Ariel} can achieve its scientific requirements in the presence of the expected pointing jitter. The requirements (for bright targets) are expressed as\footnote{European Cooperation for Space Standarization (2008). Control Performance Standard ECSS-E-ST-60-10C. ESA-ESTEC Requirements \& Standards Division.}\textsuperscript{,}\footnote{ESA Engineering Standardization Board (2011). ESA pointing error engineering handbook ESSB-HB-E-003. European Space Agency (ESA).} drifts smaller than 70\,mas at timescales longer than 90\,s (Performance Drift Error, PDE $<$ 70\,mas from 90\,s to 10\,hr), an RPE (Relative Pointing Error) smaller than 130\,mas at time scales between 0.1\,s and 90\,s, and an RPE smaller than 180\,mas at timescales shorter than 0.1\,s. All quantities are intended at a confidence level of 99.7\%. 

The \textit{Ariel} detectors are Teledyne's HxRG Mercury-Cadmium-Telluride (MCT) that are read up-the-ramp in a sequence of Non-Destructive Reads (NDRs) before the detector is reset to prevent saturation. The readout pattern is known as MULTIACCUM (see e.g. \cite{Rauscher:2007}), where each group consists of a single NDR. Therefore, previous analyses consider both an ``instantaneous readout,'' where the focal plane array is ``read'' in an instantaneous snapshot as if there were a shutter, and a ``sequential readout,'' which mimics the actual readout process where pixels are read line-by-line with a typical pixel cadence of 10 \textmu s in Ch0, and 8 \textmu s in FGS-1.

A Correlated Double Sample (CDS) frame, or CDS science frame, is estimated from the difference between the last and the first NDR in each exposure. This process nulls the kTC noise term and minimizes the noise for shot-noise-limited observations \cite{Rauscher:2007}. Each CDS science frame is then flat-field calibrated, assuming flat-field coefficients are known with a moderate 0.5\% precision with respect to their true value. The angular shifts in pitch and yaw among CDS science frames are estimated by cross-correlating each frame with a reference frame (e.g., the first or the median across the entire simulated observation). Each CDS frame is subsequently registered on a common grid: because all focal planes are Nyquist sampled, this operation can be performed without introducing artifacts. Spectro-photometric timelines are then obtained using aperture photometry.

While these simple reduction steps demonstrate compliance with the jitter-detrending requirements, their performance is limited: (i) they are not always capable of reducing the residual noise below the required 20~ppm at timescales longer than 1~hour in the presence of jitter with non-stationary statistical properties, of the type described in the next section, and (ii) they rely on the accurate knowledge of flat-field coefficients.

Photometric noise caused by jitter arises from the non-uniformities of the focal plane. Photon Response Non-Uniformity (PRNU) models the variation of Quantum Efficiency (QE) from pixel to pixel. Additionally, the surface area of a pixel is not uniformly responsive to incident photons, and the efficiency of photon detection across a pixel surface is quantified by the Intra-Pixel Response Function (IPRF). Due to its Nyquist-sampled focal planes, \textit{Ariel} is insensitive to the IPRF under linear shifts of focal plane images. The effect of jitter can be thought of as a blurring of the CDS image, i.e., a convolution of the CDS image's true value with a jitter kernel, and noise. If the jitter kernel is stationary in shape, flat fielding can mitigate the systematic arbitrarily well by increasing the precision of the flat field calibration, despite possible drifts in the line of sight (LoS).

However, the statistical properties of jitter, e.g., the Root Mean Square (RMS), are not stationary, and the jitter convolution kernel varies over time. The net result is that CDS signals are sampled by slightly different pixels from time to time ($\ll 1/10$\,pixel), even in the absence of drifts. All of this results in a photometric systematic that is well correlated with the properties of the jitter convolution kernel, which can in turn be inferred from the variations in the shape (i.e., light distribution) of the CDS image frames.

A vast literature reports several examples of successful detrending of time-correlated measurements of exoplanet spectra, including from the Hubble and Spitzer space telescopes \citep[e.g.][]{Charbonneau_2005, Charbonneau_2008, Knutson_2009, Ballard_2010, Stevenson_2012, Gibson_2011, Deming_2015, Morello_2014, Morello_2015, Swain:2008, Schwartz_2016}. Among these, \citet{Ingalls:2016} provide a summary studying the repeatability of atmospheric spectra extraction using different detrending methods for post-cryogenic Spitzer/IRAC. Developed as part of the Spitzer 2015 data challenge, these methods\footnote{Notably BLISS mapping, Pixel Level Decorrelation, and Independent Component Analysis.} leveraged the spatial information in the science frames to detrend time-series IRAC observations affected by jitter, PRNU, and IPRF.

In this work, we aim to enhance the data reduction and jitter-detrending capabilities of the \textit{Ariel} mission demonstrated so far. Specifically, we seek to design a jitter detrending algorithm capable of minimizing jitter disturbances relative to the photon noise limit at all timescales (beyond 1 hour), even in the presence of non-stationary jitter, and without utilizing flat-field calibration. We opt for an approach aligned with IRAC's, utilizing spatial information on the Point Spread Function (PSF) distortion from jitter to detrend the optical signals. For simplicity, we focus our study on one photometer (FGS-1) and one spectrometer (AIRS Ch0),  as extension to other focal planes is straightforward.

\subsection{Data generation}\label{sec:methods:data-generation}

The simulations for this study are produced using \texttt{ExoSim2}, a generic time-domain simulator of spectro-photometric observations of transiting exoplanets developed by the AMC. Using a parametric description of the payload, \texttt{ExoSim2} can simulate an entire \textit{Ariel} observation as a function of time, enabling detailed performance studies during all mission phases. 

The simulated observations consist of photometric and spectral images vs. time, sampled in NDR up-the-ramp. A series of NDRs is acquired during each exposure, from which CDS science frames are estimated from the difference between the last and the first NDR in each exposure. This maximizes the Signal-to-Noise Ratio (SNR) for the \textit{Ariel} observations dominated by the target host star photon noise. 

Instrument effects included in the simulation are: i) photon noise from the star, ii) detector dark current and readout noise (20 e$^-$/CDS/pixel), iii) jitter, iv) 5\% PRNU, v) IPRF with parameters form \citet{Barron_2007} and model by  \citet{Pascale:2015} (their Eq.~12). This preliminary study excludes other effects, as they are expected to have a negligible impact on the reduction quality. These effects will be accounted for in future studies. 

The simulations utilize PSF models based on the optical design of the \textit{Ariel} FGS-1 and AIRS Ch0. The PSFs are estimated using \texttt{PAOS}~\citep{Bocchieri:2024}, a generic Physical Optics Propagation tool developed for \textit{Ariel} and publicly available as open-source software\footnote{\url{https://paos.readthedocs.io/}}. 
\textit{Ariel} is diffraction-limited at wavelengths above 3~\textmu m and relies on geometric aberrations to ensure that shorter wavelength focal planes are critically sampled, including FGS-1. The level of geometric aberrations introduced can be controlled by the actuated telescope secondary mirror that can add or remove defocus to the propagating wavefront. Therefore, we use PAOS to estimate diffraction-limited PSFs for AIRS Ch0 and consider a PSF with a level of geometric aberrations and defocus to comply with the Nyquist sampling requirement at FGS-1. It is noteworthy that the level of aberrations needed to ensure that the FGS-1 focal plane is critically sampled is small when compared to the wavelength of Ch0, which is negligibly affected. In future work, we will extend the analysis to use the same wavefront error across all focal planes, although it does not impact the results.

The jitter model used is provided by ADS in the form of a simulated time series of pitch and yaw of the boresight spanning 10 hours and sampled at 1 kHz. The simulation is expected to be a representative realization and includes: (i) vibrations exported by the payload cryocooler required by the AIRS focal plane arrays; (ii) vibrations exported by the accelerating RWs; (iii) uncertainty on the centroid estimate of the FGS guiding channel; and (iv) possible friction jumps in the RW ball bearings. The simulated time series also includes a slow linear angular drift ($\simeq 2$ mas/hr).
This last effect was removed before injecting the time series into the simulator because (i) it is expected to have no negative impact on the detrending (because it is deterministic), and (ii) it will likely not be present, or present with a significantly smaller amplitude.

\begin{figure}[htbp]
    \centering
    \includegraphics[width=1\linewidth]{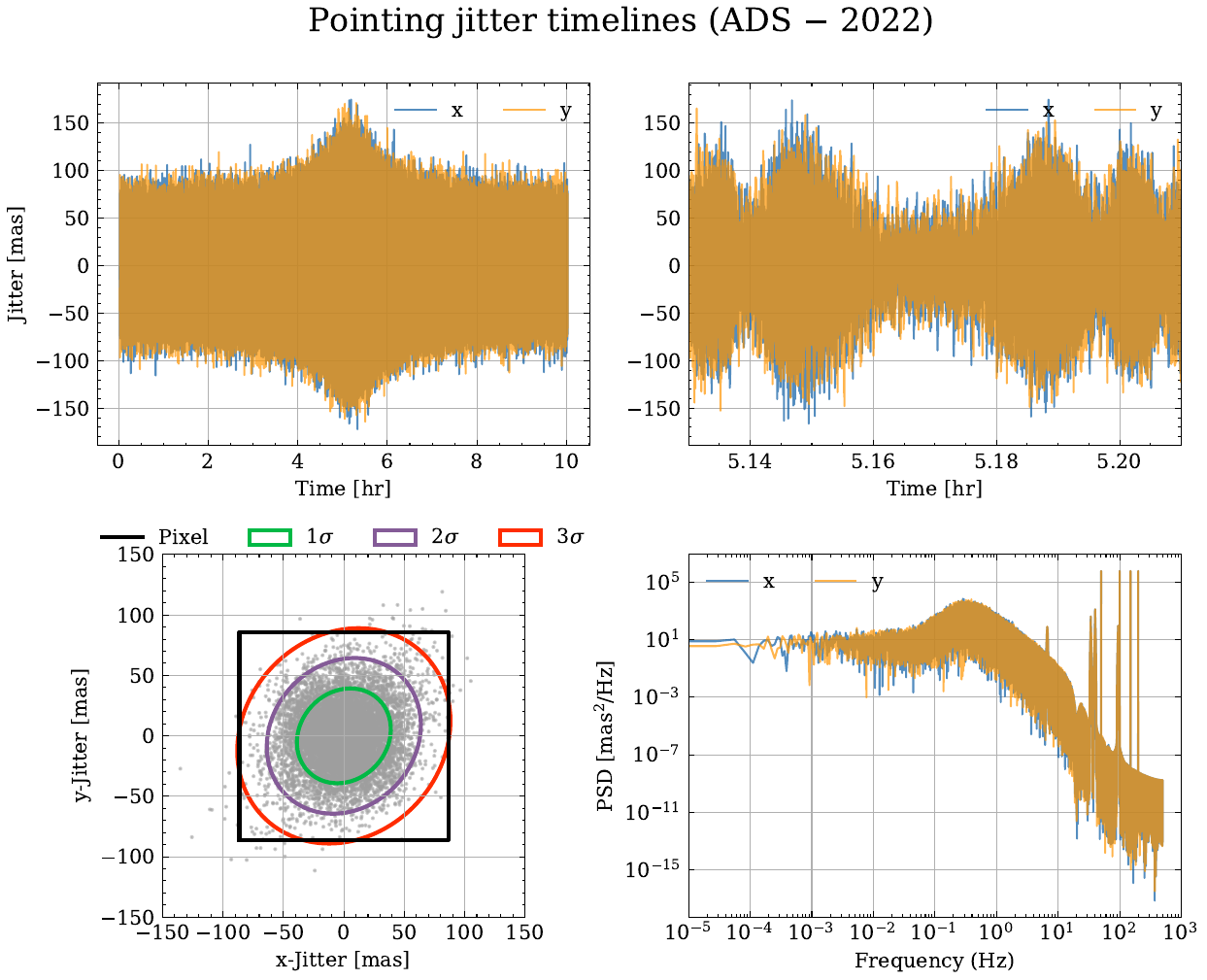}
    \caption{
    Simulated jitter time series provided by ADS. Rotations around the $x$ and $y$ axes are respectively yaw and pitch.   
    \textit{Top left}: The extended, 10-hour timelines illustrate the slow change in the jitter RMS.
    \textit{Top right}: A version of the left plot, zoomed in around the center of the timelines, showcasing the non-stationarity of the RMS at timescales of minutes.
    \textit{Bottom left}: Scatter plot of the jitter (``jitter ball''). 
    The size of a pixel in FGS-1 is shown for reference. 
    Contour ellipses indicate the jitter distribution's 1-, 2-, and 3-$\sigma$ confidence intervals, estimated from the covariance matrix.
    \textit{Bottom right}: Power spectral densities.}
    \label{fig:jitter-timeline}
\end{figure}
Figure~\ref{fig:jitter-timeline} illustrates the jitter time series corresponding to the nominal case for a bright target, showing a non-stationary RMS, as mentioned earlier. In this particular realization, the jitter RMS exhibits a slow rise in amplitude from the beginning to about the center of the time series, followed by a corresponding decrease in amplitude from the center to the end of the timeline (see top left). The envelope of the time series ranges from -150 mas to 150 mas, and the jitter RMS is less than approximately 50 mas (see bottom left)\footnote{The scatter plot contains only 10,000 randomly drawn points to accommodate image size.}. A closer inspection of a portion of the time series near the center (see top right) reveals additional structure in the RMS at timescales of minutes.

Overall, the jitter exhibits non-white power spectral densities across a wide frequency range, with narrow harmonic lines corresponding to the exported vibrations from the cryocooler (see bottom right). Jitter, however, is white at frequencies smaller than approximately 100 mHz. As a consequence, the jitter-induced photometric noise is non-white: if left uncorrected, it would dominate the noise content of the photometric and spectroscopic timelines, biasing the estimates of the exoplanet spectrum.

The simulation generates a 10-hour-long mock observation of a star similar to HD~209458. With a K-band magnitude of $\simeq 6.3$ it is a bright target for \textit{Ariel}, chosen because jitter effects are stronger for brighter targets. 
Instead of using more realistic stellar models, we simulate the star's spectral irradiance using a Planck function estimated at the star's effective temperature. This choice was initially made because no significant differences are expected between using synthetic stellar spectra and a Planck function, and we preferred using a well-behaved spectral dependence for this preliminary study. The simulations can easily incorporate more realistic spectral irradiance and stellar activity models in future studies, as ExoSim2 already has these capabilities implemented. The exposure times are set to 3 seconds to avoid detector saturation.

The observation is further modulated with the transit of a planet similar to HD\,209458b using \texttt{batman}~\citep{Kreidberg:2015}, implementing a flat transmission spectrum and no stellar limb-darkening, to ease the interpretation of the results. The mid-transit time is placed exactly at hour 5 in the ADS jitter simulation. We have empirically verified that this is the worst case compared to a mid-transit time shifted with respect to the mid-time in the jitter time series.

In addition to simulations with jitter, we also generate simulations without jitter. These jitter-free simulations are affected only by unavoidable uncertainties such as photon noise and detector noise. The reduction of jitter-free simulations provides a \textit{reference} for comparing the reduction process when jitter is present. The effectiveness of the jitter detrending algorithm is indicated by how closely the noise level of the jitter-affected observations matches the noise level of the jitter-free observations. The closer the noise levels, the more effective the jitter detrending.

\subsection{Signal extraction}\label{sec:methods:signal-extraction}

The signal extraction process begins with the CDS science frames, which are processed similarly for both the photometer and the spectrometer, albeit with some notable distinctions outlined below. FGS-1 CDS frames have dimensions of 64x64 pixels, while Ch0 frames measure 64x354 pixels.

For spectroscopic CDS images, the relative positional shift along the dispersion and cross-dispersion directions is estimated by cross-correlating each frame with a reference, such as the median of all CDS frames out of transit or the first CDS image in the observation. The choice of reference frame does not impact the results. Subsequently, all spectroscopic CDS images are registered onto a common pixel grid by applying the positional shifts (changed in sign) estimated from the cross-correlation. The registration is done in Fourier space for sub-pixel precision \citep{Guizar-Sicairos:2008}. This registration process ensures no artifacts are introduced, given that the \textit{Ariel} spectroscopic focal planes are critically sampled. This step aligns with the jitter detrending approach used in previous mission phases, as discussed earlier in this section. The photometric CDS frames remain unshifted, as this does not negatively impact the precision of the photometry.

According to science requirements, both photometric and spectroscopic timelines must be sampled with a cadence faster than 90 seconds to capture the ingress and egress events of the light curves. Consequently, we assume a cadence of 30 seconds, achieved by summing 10 consecutive CDS frames acquired with a 3-second cadence. This operation not only enhances the SNR of both photometric and spectroscopic images (particularly at the red end of Ch0), but also averages down the effects of jitter. Although not strictly necessary, the higher SNR of the binned photometric and spectroscopic images simplifies the analysis, as working with low-noise data reduces complexity. Since the dispersion axis of the spectrometer aligns with the detector pixel rows, the spectral images are binned in wavelength by summing three consecutive pixels in each row, resulting in spectral image dimensions of 64x118 pixels after binning.

\begin{figure}[htbp]
    \centering
    \includegraphics[width=0.425\textwidth]{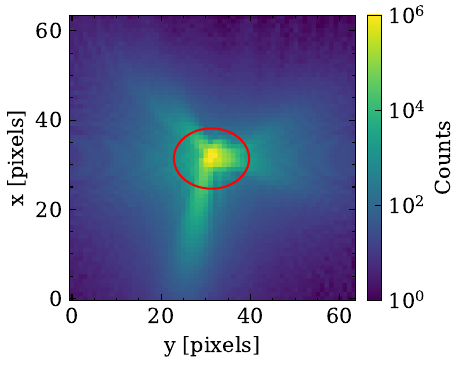}
    \includegraphics[width=0.565\textwidth]{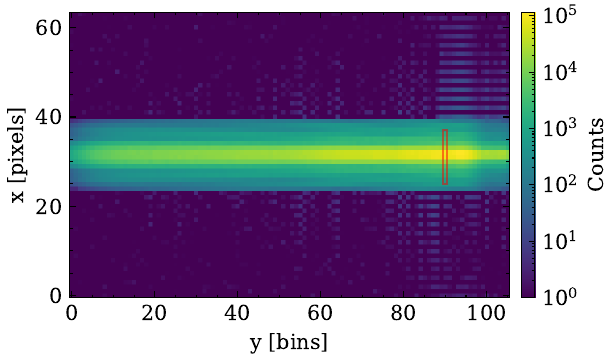}
    \caption{
    \textit{Left}: Median CDS-science frame taken out of transit and photometric aperture of FGS-1. 
    \textit{Right}: Same, for AIRS-Ch0. 
    For better visualization, we draw the aperture for a single spectral bin.
    }
    \label{fig:apertures}
\end{figure}

The signal extraction process employs aperture photometry, implemented using \texttt{photutils}\footnote{\url{https://photutils.readthedocs.io/}.}. Figure~\ref{fig:apertures} provides a visual representation of typical CDS science frames and the photometric apertures utilized in the analysis. The FGS1 focal plane aperture is an ellipse centered on the PSF centroid (described in the next section). The orientation and ellipticity of the ellipse correspond to the shape of the telescope's input pupil, which measures 1.1\,m x 0.73\,m based on requirements. The dimension of the aperture is sized on the median photometric image out of transit, such as to contain 95\% of the signal. Given that the wavelength axis of the AIRS Ch0 spectral images has already been binned, Ch0 photometry is extracted using a rectangular aperture that is 12 pixels tall and 1 pixel wide, fixed around the center of the optical signal registered on a common grid. Both the photometric and spectroscopic apertures are optimized to empirically maximize the signal-to-noise ratio (SNR) for these observations limited by the photon noise of the target.

Finally, the photometric and spectroscopic light curves are normalized to unity out-of-transit. This step is not strictly necessary, but it is implemented for convenience, as it allows for working with signal values close to unity.

It is important to highlight that in this work, no calibration products are utilized to process the science frames. This includes flat-fielding, a process where the detector's pixel response is normalized using an estimate of the QE matrix, which represents the response to uniform illumination. Flat-fielding is typically part of the \textit{Ariel} baseline reduction and has been employed in previous analyses. However, experiences with instruments like HST, Spitzer and now JWST suggest that auto-calibration of science frames can be effective for exoplanet observations. Auto-calibration adds redundancy to the data reduction process and avoids the need for calibration measurements and their associated biases. Consequently, our methodology aims to auto-calibrate \textit{Ariel} data by fully leveraging the spatial information present in the CDS-science frames.

\subsection{Jitter-detrending algorithm}\label{sec:methods:algorithm}

The photometric noise introduced by jitter, as explained earlier in this section, operates through a specific mechanism: the displacements (``shifts'') and changes of shape (``distortions'') in the PSF result in the optical signal being sampled by slightly different focal plane pixels. Since these pixels have varying responses to the incoming optical power, it leads to fluctuations in the sampled signal. Consequently, these sampled signal variations must be correlated with the shifts and distortions in the PSF. 

The mathematical model describing the sampled signal vs. the incident signal can be written as
\begin{equation}
    S_\lambda(t) = f_\lambda(\vec{X}) \times \phi_\lambda + n_\lambda(t)
    \label{eq:signal}
\end{equation}
In this equation,  $S_\lambda(t)$, $\phi_\lambda$, and $n_\lambda(t)$ represent the sampled signal, the incoming optical signal, and the noise, respectively,  at wavelength $\lambda$ and time $t$. The jitter nuisance is expressed by the unknown wavelength-dependent function $f_\lambda$ of the parameters $\vec{X}$. These parameters model the position and distortions vs. time of the light distribution on the FGS1 and Ch0 focal planes, illustrated in Figure~\ref{fig:apertures}.

The PSF distortion parameters $\vec{X}$ are determined from the low-order moments of the light distribution on the focal plane in each photometric and spectral image. In each exposure, the position component of $\vec{X}$ represents the average shift of the LoS of the telescope ($x$ and $y$), while the width reflects the spread of the PSF ($w_x$ and $w_y$). Additionally, skewness ($s_x$ and $s_y$) and kurtosis ($k_x$ and $k_y$) track the asymmetry and tails of the distribution, respectively, making them sensitive indicators of PSF distortion. For illustrative purposes, Figure~\ref{fig:shape-deformation} in the Appendix shows how the PSF distortion affects the moments.

For sufficiently small jitter, we can approximate the jitter disturbance with a linear combination of the moments as independent variables with coefficients $c_k$:
\begin{equation}
f_\lambda = 
\begin{cases}
    c_0 + c_1 x + c_2 w_x + c_3 s_x + c_4 k_x + c_5 y + c_6 w_y + c_7 s_y + c_8 k_y & \text{[Photom.]} \\
    c_0 + c_1 x + c_2 w_x + c_3 s_x + c_4 k_x & \text{[Spectrom.]}
    \label{eq:jitter-model}
\end{cases}
\end{equation}
Here, we dropped the explicit time dependencies. The coefficients $c_k$ are wavelength-dependent but do not vary with time. Although $\vec{X}$ is inherently time-dependent,  it does not depend on wavelength, as the jitter is fully correlated in all spectral bins and focal planes. 

Therefore, the jitter-detrending algorithm is implemented as follows:
\begin{enumerate}
    \item Estimate position, width, skewness, and kurtosis from the science frames;
    \item Register all spectral images on the same grid using the estimated first-order moments ($x$ and $y$), but leave all photometric images unshifted;
    \item Simultaneously\footnote{The coefficients can not be estimated independently as the moments are not statistically independent stochastic variables.} correlate the estimated parameters with the measured signal;
    \item Remove the identified correlations from the measured signal.
\end{enumerate}

In practice, when the observed light curve contains the signal of a transiting planet, the model  
\begin{equation}
    S_\lambda(t) = f_\lambda(\vec{X}) \times \phi_\lambda \times \Lambda_\lambda(\vec{P}, t) + n_\lambda(t)
    \label{eq:signal2} 
\end{equation}
is extended to include the transiting light curve  $\Lambda_\lambda(\vec{P}, t)$ \citep{mandel_agol:2002}. Here,   $\vec{P} = \{R_p/R_s,  a/R_s, i, e, t_0, P\}$ represents the exoplanet parameters: the planet's radius relative to the star's radius, $R_s$, the orbital semi-major axis in units of stellar radii, orbital inclination, eccentricity, time at mid-transit, and period, respectively. The exoplanet parameters $\vec{P}$ and the nuisance parameters $c_k$ are estimated simultaneously by fitting the above model to each photometric and spectroscopic timeline.  

The low-order moments of the focal plane light distribution for FGS-1 are estimated for two orthogonal cross sections along the pixel rows and columns of the focal plane images. These cross sections are centered around an estimate of the centroid, which is obtained using a light barycenter method. Because jitter is sub-pixel, the same row and column are used consistently throughout the entire observation. We estimate \(x, w_x, s_x, k_x\) from the vertical section and \(y, w_y, s_y, k_y\) from the horizontal one. Here $x$ and $y$ represent centroids with improved precision upon the first estimate. 

These improved centroids are estimated using the Iterative Weighted Center of Gravity (IWCoG) algorithm \citep{akondi:2010, akondi:2013}, which is a modified version of the simpler light barycenter algorithm. In the IWCoG, the image is first multiplied with a weighting function, which is similar in shape to the PSF and centered on the current centroid estimate. After determining the initial centroid, the weighting function is placed at the calculated position, and the centroid is determined again. This process is repeated multiple times with increasing precision until convergence is achieved. 

The weighting function used is a Gaussian that is  $2\sqrt{2}$ wider than the best Gaussian fit to the PSF cross sections. This widening factor has been empirically found to maximize the correlation between the moments and the signal and can be adjusted in the future if needed. Once the high-precision centroid estimate has converged, it is used to center the Gaussian weighting function for the estimation of higher moments (width, skewness, and kurtosis).

The moments of each spectral bin along the cross-dispersion direction in the spectroscopic timelines are estimated similarly, with the following main differences: i) the centroid estimate is replaced by the estimate of the linear shifts obtained from cross-correlation; ii) the weighting function used is a Gaussian that is  $\sqrt{2}$ wider than the best Gaussian fit to the PSF cross sections. As explained earlier, the jitter disturbance is correlated across all spectral bins, so the moments are estimated from a portion of the spectroscopic frames where the SNR is largest. For this study, we use the five brightest spectral bins to estimate the moments and jitter-detrend the spectroscopic timelines. A future study may consider a weighted estimate of the moments across the spectroscopic images.

The observed photometric and spectroscopic light curves are then fit with the model of Equation \ref{eq:signal2}. This fitting is performed with a non-linear least squares method, implemented via the \texttt{curve\_fit} function from the \texttt{scipy.optimize} library. The free parameters are $c_k, R_p/R_s, a/R_s, i, t_0$, with all other parameters fixed to their true values. A stellar model with undarkened limbs is used. Using Monte Carlo methods to estimate the model parameters provides equivalent results.

\subsection{Noise analysis}\label{subsec:noise-analysis}

The noise in the residual photometric and spectroscopic light curves, after subtracting the best-fit model, is analyzed for three different cases:

\begin{enumerate}
    \item \textit{raw}: the signal obtained directly from aperture photometry (no detrending);
    \item \textit{detrended}: the signal after jitter detrending;
    \item \textit{reference}: the signal timeline without any jitter.
\end{enumerate}

We use the Allan deviation (ADEV)~\citep{Allan:1987} in its overlapping form as the noise metric. The Allan deviation is a statistical tool designed to evaluate the amplitude of the noise in time-series data at different time scales. ADEV curves can also serve as a basis to quantify the prevalence of different noise contributions (e.g., white noise, pink noise, Brown noise, etc.), although this is beyond the scope of this work. Instead, we focus on the noise level achieved post-processing, given the implemented jitter-detrending algorithm.
See Appendix~\ref{appendix:stability} for details. 

Our simulations consider only jitter and random noise, specifically photon noise from the star and detector readout noise. The jitter is represented by a single realization provided by ADS. To avoid biases from a single realization of random noise, we perform our analysis on 128 independent noise realizations, all using the same jitter timeline. This approach allows us to estimate uncertainties in the parameters and assess whether the jitter-detrending algorithm introduces any statistical bias.

Using multiple jitter timeline realizations would increase the robustness of our results. We plan to repeat the assessment as new jitter realizations become available. However, we do not expect different results using different jitter realizations, as the realization used in this work is expected to be representative. In the past, we have tested our detrending methods on less representative, but significantly worse, jitter models to stress-test our detrending algorithms with similar results to those reported in this work \citep{Tinetti:2021}.

\section{Results}\label{sec:results}

A data reduction pipeline is applied to the simulated data to produce the light curves and spectra of the target. 
The pipeline includes all the steps described in the previous Section~\ref{sec:methods}, from signal extraction and moments estimation to jitter detrending.

An optimized data reduction pipeline mitigates systematic uncertainties without introducing biases. In the ideal post-processed signal, any residual systematics should be sub-dominant compared to the fundamental noise sources of target photon noise and detector noise.

To assess the results of our reduction pipeline, we divide this section into two parts: Section~\ref{sec:results:noise} focuses on the noise in the post-processed signal, and Section~\ref{sec:results:bias} investigates the biases in the retrieved atmospheric spectrum.

\subsection{Noise}\label{sec:results:noise}

\begin{figure}[htbp]
    \centering
    \includegraphics[width=1\linewidth]{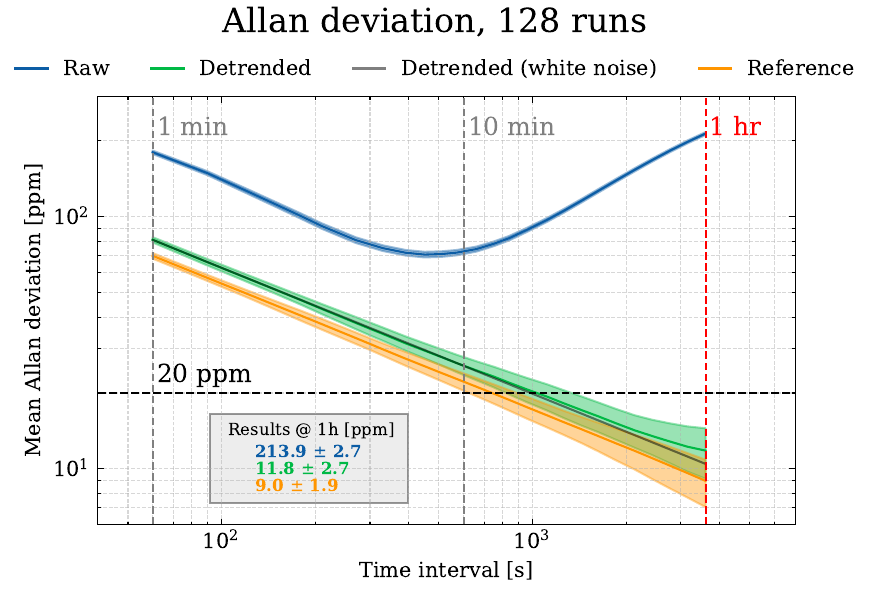}
    \caption{
        Stability analysis of the FGS-1 photometer, illustrating the overlapping Allan deviation (ADEV) of the residuals between the obtained light curve and the ground truth model. 
        The mean ADEV of the \textit{raw} (blue), \textit{detrended} (green), and \textit{reference} (orange) residuals are shown. The mean is taken over 128 noise realizations, using the same jitter timeline. 
        The 1-$\sigma$ deviation around the mean is shown as a shaded area. 
        The ADEV is computed for different time intervals and is expressed in parts-per-million (ppm). 
        Vertical dashed lines indicate relevant time scales for the analysis. 
        The gray line represents the expected ADEV of the \textit{detrended} residuals if it were random white noise. 
        The text box at the bottom right corner shows the mean ADEV and associated 1-$\sigma$ deviation at the 1-hour timescale for the residuals, with the same color coding.
    }
    \label{fig:oadev-residuals-n}
\end{figure}

Figure~\ref{fig:oadev-residuals-n} shows the ADEV of the residuals vs.~integration time interval, $\tau$, for the photometer. The simulated timeline spans a representative duration of 10 hours, and the ADEV is estimated up to $\tau = 1$\,hr to ensure adequate sampling of the variance at this time scale.

Random time-uncorrelated noise processes dominate the \textit{raw} residual timeline up to a few minutes as the noise integrates down roughly as $1/\sqrt{\tau}$. However, at longer time scales the Allan deviation curve bends at $\tau = 10$~min, eventually rising to exceed the initial value observed at $\tau = 1$\,min, as $\tau$ approaches 1\,hr.

In contrast, the \textit{detrended} curve exhibits the expected integration behavior for a random, time-uncorrelated process, similar to the \textit{reference} curve, albeit showing a slight excess of approximately $15$\% to 30\% compared to the latter. An apparent deviation in slope in the \textit{detrended} ADEV occurs around 2000\,s, as visible in Figure~\ref{fig:oadev-residuals-n}. This deviation may not be statistically significant, as indicated by the dispersion across the 128 realizations (semi-transparent green region in the Figure). Alternatively, it may indicate this time scale's relevance for residual correlated noise in the timeline. Further investigation would be necessary to address this question fully, although it may not be justified due to the significantly low noise level obtained ($<$ 20\,ppm), which exceeds requirements (Section \ref{sec:methods}). Still, if the interpretation of a small residual correlation is correct, then it may be possible to improve the detrending in future work. We can, however, rule out expected improvements using a flat field calibration because results with or without the flat field calibration were comparable when tested on these simulations. The impact on the estimate of the planet radius, $R_p$, is discussed in the next section.

Comparing \textit{raw} and \textit{detrended} ADEVs, we observe a significant improvement in both relative offset and slope at short and long $\tau$, respectively. As detailed in the text box in Figure \ref{fig:oadev-residuals-n}, the \textit{raw}  ADEV at $\tau$ = 1\,hr is approximately 215\,ppm, whereas the \textit{detrended}  ADEV is around 12\,ppm, representing an 18-fold reduction in noise. Meanwhile, the \textit{reference} ADEV is approximately 9\,ppm at the same timescale. Consequently, the jitter-detrending algorithm has effectively corrected most of the photometric noise introduced by the jitter, with only a small residual excess on an otherwise small noise level, when compared with requirements (20 ppm at the time scale of 1\,hr, Section \ref{sec:methods}). Given the 1-$\sigma$ deviation of the ADEV at $\tau$ = 1\,h, the stability requirement of 20\,ppm at $\tau$ = 1\,h is achieved with a 3-$\sigma$ confidence for the \textit{detrended} curve.

\begin{figure}[htbp]
    \centering
    \includegraphics[width=1\linewidth]{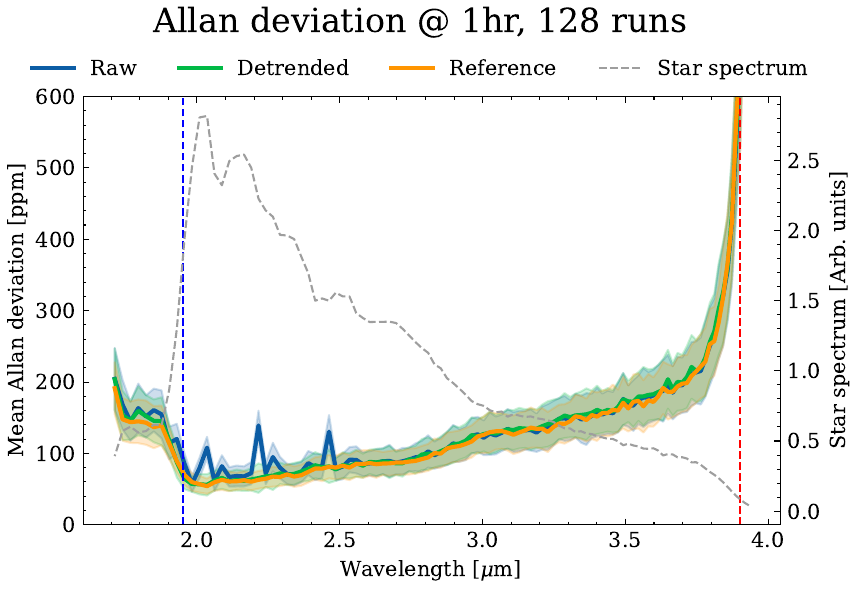}
    \caption{
        Stability analysis of the AIRS-Ch0 spectrometer, illustrating the overlapping Allan deviation (ADEV) of the residuals between the obtained light curve vs.~wavelength and the ground truth model. The mean ADEV of the \textit{raw} (blue), \textit{detrended} (green), and \textit{reference} (orange) residuals are shown. 
        The mean is taken over 128 noise realizations, using the same jitter timeline. 
        The 1-$\sigma$ deviation around the mean is shown as a shaded area. 
        The ADEV is computed at the 1-hour timescale and is expressed in parts-per-million (ppm). 
        The stellar spectrum convolved with the instrument response function is shown as a dashed gray curve for reference; its units are arbitrary. 
        The vertical dashed lines mark the ends of the nominal spectral range of the instrument, colored in blue and red, respectively.
    }
    \label{fig:mean-oadev}
\end{figure}

Figure~\ref{fig:mean-oadev} shows the ADEV of the residuals vs.~wavelength for the spectrometer, at $\tau$ = 1~h. 
Like the photometer, we show the ADEV of the \textit{raw}, \textit{detrended}, and \textit{reference} residuals, averaged over 128 noise realizations, along with the dispersion across the 128. We also provide the stellar spectrum as measured by the spectrometer to facilitate comparison between the ADEV and the signal level.

The \textit{raw} ADEV significantly exceeds the \textit{reference} between 2.0 and 2.5~\textmu m, over a spectral region where the stellar signal is most prominent. This discrepancy arises because the photon and detector noise integrate below the level of the jitter contribution. Left uncorrected, the jitter contribution remains time-correlated and does not integrate down as efficiently. Photon noise becomes more important at longer wavelengths in the noise budget, dwarfing the jitter noise, as indicated by the \textit{raw} ADEV tacking the \textit{reference}. This may suggest that jitter detrending may not be necessary when photon noise dominates the noise budget, as would be the case for targets fainter than the K-band 6.3 magnitude stellar target used in this study. However, biases in the retrieved parameters may persist, as discussed in the next section. 

Conversely, the stability of the \textit{detrended} curve indicates that the jitter-detrending algorithm effectively removes jitter-induced variability at all wavelengths and at the required timescale,  $\tau$ = 1~h. There is no excess similar to that observed for the photometer, likely due to the larger photon noise contribution in the spectrometer compared to the photometer.

These considerations further justify the choice of estimating the low-order moments used in the detrended algorithm from the blue end of the spectrum, as the signal-to-noise ratio is smaller at the red end of the spectrum.

\subsection{Bias}\label{sec:results:bias}

\begin{figure}[htbp]
    \centering
    \includegraphics[width=1\linewidth]{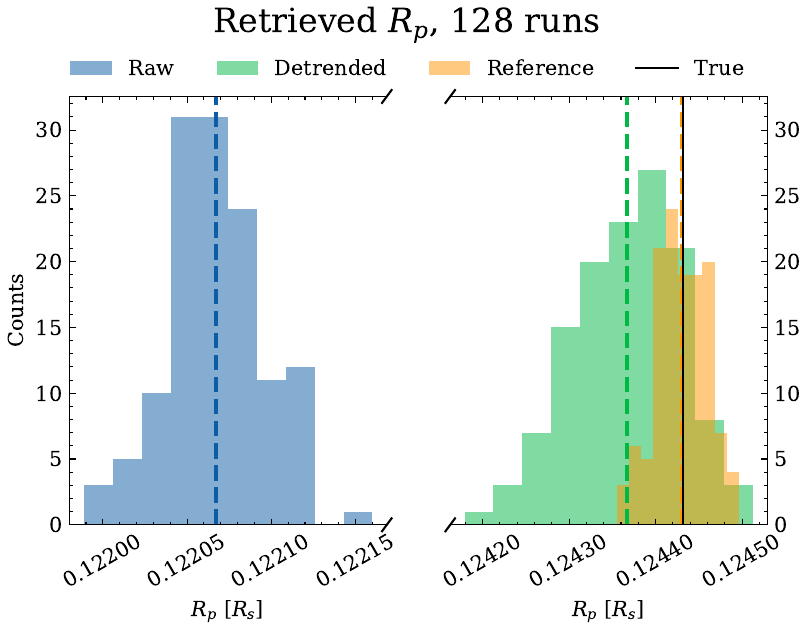}
    \caption{
        Retrieved planetary radius for the FGS-1 photometer. 
        The histograms show the distribution of the retrieved planetary radius in units of stellar radius, for the \textit{raw} (blue), \textit{detrended} (green), and \textit{reference} (orange) light curves. 
        The histograms are created using 128 noise realizations, using the same jitter timeline. 
        Note the break in the x-axis. 
        The solid black line represents the ground truth value, while the dashed lines are the mean of the retrieved values of each histogram.
    }
    \label{fig:rp-over-rs-hist}
\end{figure}

To investigate potential biases in the retrieved science signal parameters, we focus on the planetary radius, \(R_p\), measured from the \textit{raw}, \textit{detrended}, and \textit{reference} light curves. This analysis uses the same 128 noise realizations discussed in the noise analysis section, i.e., simulations repeated for 128 different noise realizations, but using the same jitter time series. Figure~\ref{fig:rp-over-rs-hist} shows the histograms of the retrieved planetary radius for the photometer in the three cases.

The distribution of $R_p$  for the \textit{reference} is unbiased, meaning the expected value of the planet radius coincides with the true value used in the simulation. The dispersion of approximately 30 ppm on $R_p$ results in an estimated uncertainty on the transmission spectrum, $\left( R_p/R_s\right)^2$, of approximately 7.2 ppm using error propagation. This is consistent with the noise estimated in the previous section of $9.0\pm 1.9$\,ppm at a timescale of one hour, for the transiting planet used in the simulation with a transit duration slightly shorter than 3 hours. 
 
The \textit{raw} histogram shows a distribution significantly biased when jitter is left uncorrected. In contrast, the \textit{detrended} distribution of retrieved $R_p$ is close to the \textit{reference}, and it is only slightly shifted from the true value. The shift, or bias, between the mean and the true value is $65 \pm 5$ ppm, and the standard deviation is approximately 60 ppm, about twice as large as the \textit{reference} histogram. A $\chi^2$ test compares the 128 $R_p$ estimates with their true values. The uncertainty for each parameter $R_p$ realization (used for the $\chi^2$ estimation) is determined starting from the RMS of the residuals of the photometric and spectroscopic light curves, and propagating it to the uncertainty on the parameter. This is implemented by setting \verb|absolute_sigma| to True in \verb|scipy.optimize.curve_fit|. The reduced $\chi^2$ obtained is 710 (\textit{raw}), 1.7 (\textit{detrended}), and 1.0 (\textit{reference}).  The large $\chi^2$ of the \textit{raw} and the unity $\chi^2$ of the \textit{reference} are expected. The \textit{detrended} case can be tentatively interpreted as follows. The random noise (60\,ppm) that is larger compared to the \textit{reference} (30\,ppm) may be caused by residual correlations in the detrended light curve residuals discussed in the previous section (also, Figure~\ref{fig:oadev-residuals-n}, yellow curve). The 65\,ppm bias may be the effect of residual systematic uncertainties, of which we are seeing only one possible realization because we are i) not randomizing the payload properties and/or ii) we have a single realization of the jitter timeline. The equivalent uncertainty on the transmission spectrum is 15\,ppm (random) and 16\,ppm (systematic). The combined effect is below the required noise floor of the payload (20\,ppm) and compliant with mission requirements. These considerations apply only to very bright targets: targets fainter than K-band magnitude $\simeq 6.3$ are likely to be dominated by their photon noise rather than the residual jitter systematics.

\begin{figure}[htbp]
    \centering
    \includegraphics[width=1\linewidth]{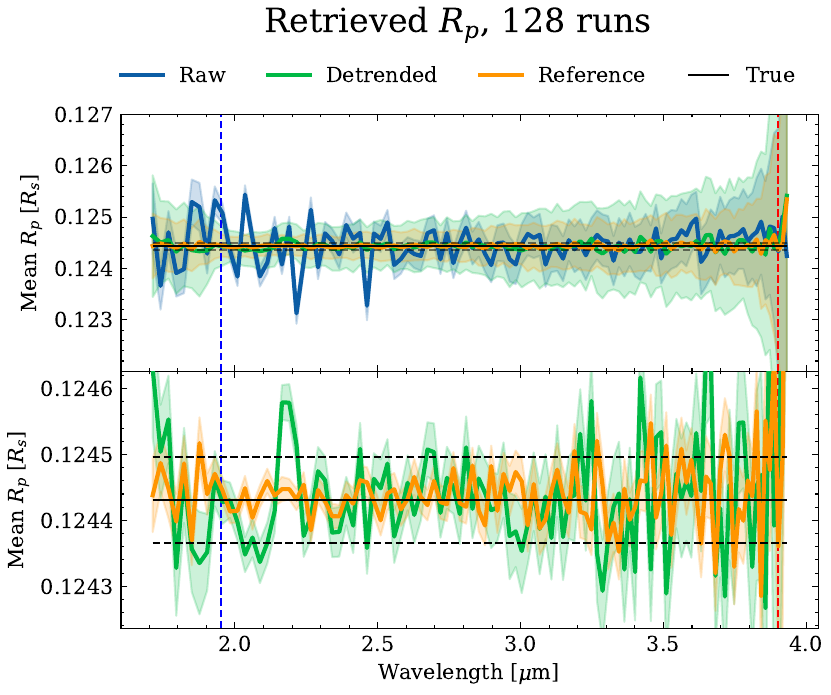}
    \caption{
        Retrieved planetary radius vs.~wavelength for the AIRS-Ch0 spectrometer. 
        For each spectral bin, the mean retrieved planetary radius in units of stellar radius is shown for the \textit{raw} (blue), \textit{detrended} (green), and \textit{reference} (orange) light curves. 
        The mean is taken over 128 independent noise realizations, but using the same jitter timeline.   The 1-$\sigma$ deviation around the mean is shown as a shaded region in the top panel. The shaded region in the bottom panel indicates the estimated uncertainty on the mean $R_p$.  A solid black line represents the $R_p$ true value, and the dashed horizontal lines indicate a band of $\pm$ 65\,ppm around it. The nominal blue and red ends of the instrument's spectral range are marked with vertical dashed lines of the same color.
    }
    \label{fig:mean-rp-over-rs}
\end{figure}

Figure~\ref{fig:mean-rp-over-rs} shows the retrieved planetary radius vs.~wavelength for the AIRS Ch0 spectrometer. In the top panel, the mean retrieved planetary radius at each spectral bin is plotted, along with the associated 1-$\sigma$ dispersion estimated from the 128 \textit{raw}, \textit{detrended}, and \textit{reference} light curves realizations. The bottom panel provides a zoomed-in view of the top panel around the y-axis, along with an estimate of the uncertainties on the mean.  

Like in the case of FGS-1, the raw curve exhibits significant excess systematic uncertainties or biases when the jitter is left uncorrected, particularly between 2 and 2.5\,\textmu m, consistent with the noise analysis showing significant excess over the \textit{reference} in this spectral region. However, significant systematics in the estimated planetary radius are observed even at longer wavelengths, despite the absence of excess noise above the \textit{reference}  in this spectral range.

The \textit{detrended} result is generally compatible with the \textit{reference}. The two horizontal dashed lines in the figure mark the $\pm$ 65\,ppm region around the true value, for comparison with the similar bias observed for FGS-1. It can therefore be seen that any excess systematics may play a role over the 2 to 2.5\,\textmu m spectral region. But, these are generally within the 65\,ppm level and therefore may comply with payload requirements. At longer wavelengths, where the stellar signal is smaller and photon noise has a larger contribution, there is no evidence of significant discrepancies between the \textit{detrended} and \textit{reference} estimates of the exoplanet radius.

\begin{figure}[htbp]
    \centering
    \includegraphics[width=1\linewidth]{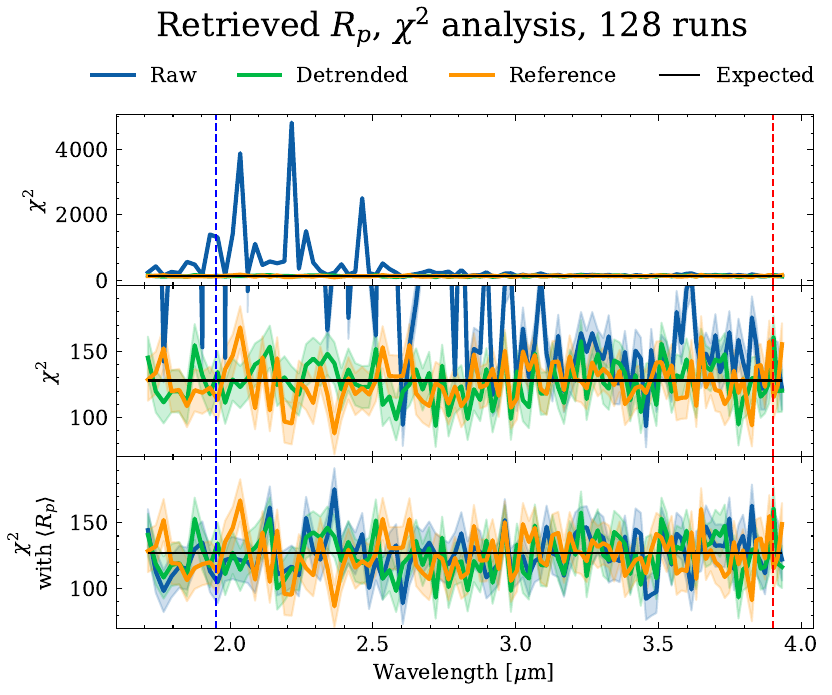}
    \caption{
        \textit{Top panel}: $\chi^2$ analysis of the AIRS-Ch0 spectrometer, illustrating the $\chi^2$ with 128 degrees of freedom (d.o.f.) between the retrieved planetary radius vs.~wavelength and the ground truth value. 
        The expected value of the $\chi^2$ is shown as a solid black line for reference. 
        The shaded areas around the \textit{raw} (blue), \textit{detrended} (green), and \textit{reference} (orange) $\chi^2$ are the expected 1-$\sigma$ deviation for a $\chi^2$ with the same d.o.f. 
        The vertical dashed lines mark the ends of the nominal spectral range of the instrument, colored in blue and red, respectively.
        \textit{Middle panel}: Same as above, but zoomed in on the y-axis to highlight the $\chi^2$ values in the range of 60 to 200.
        \textit{Bottom panel}: The $\chi^2$ here is computed using the mean retrieved value of the retrieved planetary radius vs. wavelength. 
    }
    \label{fig:mean-chi2}
\end{figure}

Figure~\ref{fig:mean-chi2}  presents the $\chi^2$ analysis of the spectrometer, divided into three panels. The top and middle panels display the $\chi^2$ values between the retrieved planetary radius and its true value versus wavelength, as done for FGS-1. The bottom panel shows the $\chi^2$ computed using the mean retrieved value, $\langle R_p \rangle$, instead of its true value. The $\chi^2$ values in the top panel serve as a measure of systematic uncertainties in the retrieved planetary radius, while those in the bottom panel indicate how well the associated uncertainties describe the data scatter. Therefore, Figure~\ref{fig:mean-chi2} provides a complementary, but equivalent representation of what is already shown in Figure~\ref{fig:mean-rp-over-rs}. The $\chi^2$ of the \textit{raw} shows excess above the expected value at all wavelengths in the top two panels, particularly in the blue, below 2.5\,\textmu m, but no excess in the bottom. Therefore,  uncorrected jitter is significantly biasing the estimate of $R_p$, but may not be affecting the ability to estimate the random noise component properly. 

Conversely, the \textit{detrended} curve in all panels shows a $\chi^2$ compatible with the expected value across the entire spectral range, with mild evidence of excess shortward of 2.5\,\textmu m, similarly to FGS-1. This can be used as evidence of excess systematics, but the effect is small, within the the payload requirements as discussed for the FGS-1 case.

\section{Conclusions}\label{sec:conclusions}

In this study, we have demonstrated the impact of noise and jitter correction on the retrieval of planetary radii from photometric and spectroscopic data. By analyzing the \textit{raw}, \textit{detrended} and \textit{reference} simulated light curves using 128 different noise realizations, we have shown that jitter noise correction significantly improves the accuracy of the retrieved planetary radii, even in the absence of any calibration.  

The \textit{reference} light curve, which represents the ideal jitter-free case, provided an unbiased distribution of the planetary radius with an estimated uncertainty consistent with previous noise estimations. The \textit{raw} data, however, showed a significant bias due to uncorrected jitter, underscoring the necessity of proper noise mitigation techniques. The \textit{detrended}  data, which included jitter corrections, resulted in a distribution close to the \textit{reference}, with only a slight bias, below payload requirements. This indicates that the applied detrending techniques effectively reduce noise and improve the reliability of the measurements. These findings highlight the critical importance of noise correction in exoplanet photometry, particularly for missions like \textit{Ariel}, where precise measurements of planetary atmospheres are essential.

We are currently working to extend the assessment presented here to include a number of relevant instrument effects, such as the presence of bad pixels, pixel glitches, and detector non-linearity, and slow drifts that might arise from electronics and other sources, across all focal planes. Additionally, we are incorporating astrophysical effects, such as synthetic stellar spectra and limb darkening, as well as addressing stellar activity. Furthermore, we plan to repeat the analysis on new jitter time series realizations from ADS. We may also explore techniques to further improve the detrending approach outlined in this work beyond payload requirements. We intend to report on our progress in a future publication.

\backmatter



\section*{Software}

Numpy~\citep{Harris:2020}, Matplotlib~\citep{Hunter:2007}, h5py~\citep{Collette:2023}, SciPy~\citep{Gommers:2024}, Allantools\footnote{\url{https://allantools.readthedocs.io/en/latest/index.html}.}

\section*{Acknowledgments}

This work was supported by the Italian Space Agency (ASI) with \textit{Ariel} grant n. 2021.5.HH.0.
We thank the \textit{Ariel} Software Simulators, Management and Documentation (S$^2$MD) Working Group for useful comments and discussions. We also thank the anonymous reviewer for their constructive comments and feedback.

\section*{Funding}
Open access funding is provided by Università degli Studi di Roma La Sapienza within the CRUI-CARE Compact Agreement.



\begin{appendices}

\section{Stability metrics}\label{appendix:stability}

To evaluate the noise content in a timeline, we use the \textit{Allan deviation} (ADEV)~\citep[e.g.\,][]{Riley:2008}. Also known as \textit{sigma-tau}, the ADEV is a metric designed to overcome the limitations of the standard deviation~\citep{Allan:1987}, having the advantage of being convergent for most noise types. 
Also, the ADEV is more informative than the standard deviation and can distinguish between different noise types, e.g.~white and colored noise.

The ADEV is usually employed to present results instead of the \textit{Allan Variance} (AVAR) as it can be directly compared to other sources of errors. 
For instance, an ADEV of 20 ppm at averaging time \(\tau = 1\) hour is the noise in the timeline at the time scale of 1 hour. 
Notably, there is no information difference between the AVAR/ADEV and a power spectral density representation: one can be transformed into the other (and vice versa) by convolving it with the appropriate transfer function~\citep{Allan:2016}.

The original AVAR has been largely superseded by its overlapping version, which improves the confidence of the resulting noise estimate. This version makes maximum use of a data set by forming all possible overlapping samples at each averaging time. 
Therefore, we refer to this form of the metric unless otherwise specified.

\section{Supplementary figures}\label{appendix:supplementary-figures}

\begin{figure}[htbp]
    \centering
    \includegraphics[width=1\linewidth]{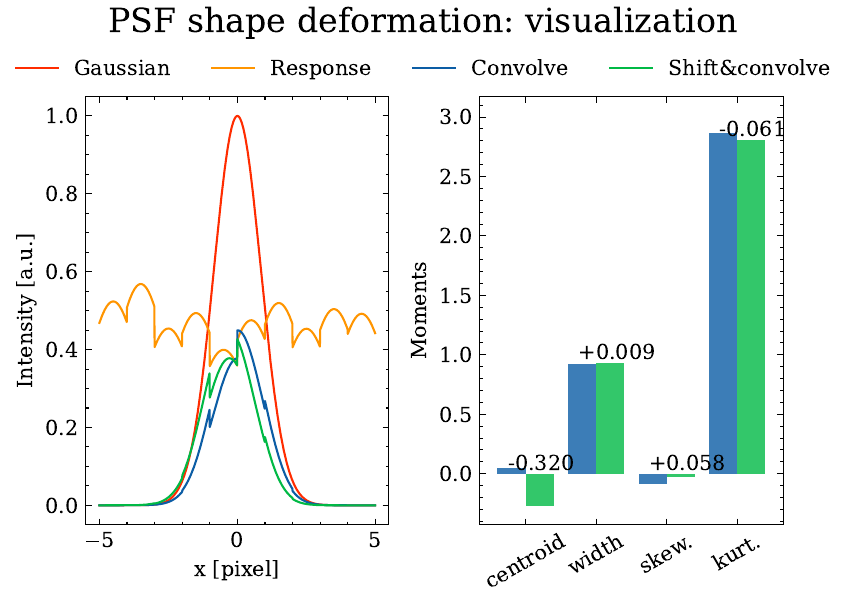}
    \caption{Visualization of the shape deformation of the PSF due to the combination of jitter movements and focal plane imperfections.~\textit{Left panel}: The PSF is represented with a Nyquist-sampled Gaussian profile (red curve). The illumination profiles before and after the jitter (sub-pixel shift of -0.3 pixels) are obtained by convolving the Gaussian profile with the pixel IPRF and multiplying by the QE. The product of the QE and IPRF is shown in the orange pixel responsivity curve for reference.~\textit{Right panel}: bar plot of the first order moments of the illumination profile before and after the jitter shift, using the same colors. The variation in the moments from the initial position is reported on top of the bars of the shifted profile.}
    \label{fig:shape-deformation}
\end{figure}

Figure~\ref{fig:shape-deformation} illustrates the mechanism of the PSF shape deformation arising from the combination of jitter movements and focal plane imperfections.

\section{Acronyms and symbols}

This section summarizes the acronyms and symbols used in the paper. The list is arranged in order of first appearance in the text.

\begin{longtable}{@{}p{3cm}p{8cm}@{}}
    \caption{Acronyms and symbols used in the text.\label{aba:tbl}} \\ 
    \hline \multicolumn{1}{|c|}{\textbf{Acronym or symbol}} & \multicolumn{1}{c|}{\textbf{Meaning}}  \\ \toprule
    \endfirsthead
    
    \multicolumn{2}{c}%
    {{\bfseries \tablename\ \thetable{} -- continued from previous page}} \\
    \hline \multicolumn{1}{|c|}{\textbf{Acronym or symbol}} & \multicolumn{1}{c|}{\textbf{Meaning}} \\ \midrule
    \endhead
    
    \hline \multicolumn{2}{|r|}{{Continued on next page}} \\ \hline
    \endfoot
    
    \bottomrule
    \endlastfoot
    \textit{Ariel} & Atmospheric Remote-Sensing Infrared Exoplanet Large-survey\\  
    ESA & European Space Agency\\
    FGS & Fine Guidance System\\
    AIRS & \textit{Ariel} Infra-Red Spectrometer\\
    S/C & Spacecraft\\
    LoS & Line of Sight\\
    AMC & \textit{Ariel} Mission Consortium\\
    ADS & Airbus Defence and Space\\
    SVM & Service module\\
    AOCS & Attitude and Orbit Control System\\
    RWs & Reaction Wheels\\
    PDE & Performance Drift Error\\
    MCT & Mercury-Cadmium-Telluride\\
    NDRs & Non-Destructive Reads\\
    RMS & Root Mean Square\\
    CDS & Correlated Double Sample\\
    PRNU & Photon Response Non-Uniformity\\
    IPRF & Intra-Pixel Response Function\\
    QE & Quantum Efficiency\\
    PSF & Point Spread Function\\
    SNR & Signal-to-Noise Ratio\\
    IWCoG & Iterative Weighted Center of Gravity\\
    ADEV & Allan deviation\\ 
\end{longtable}
    
\end{appendices}

\bibliography{biblio}

\end{document}